# Nuclear kinetic energy spectra of $D_2^+$ in intense laser field: Beyond Born Oppenheimer approximation

**Mohsen Vafaee**

Department of Chemistry, University of Isfahan, Isfahan 81746-73441, I. R. Iran

E-mail: mohsenvafaee@sci.ui.ac.ir; mohsenvafaee@gmail.com

## Abstract

Simultaneously, the vibrational nuclear dynamics and full dimensional electronic dynamics of the deuterium molecular ion exposed to the linear polarized intense laser field are studied. The time dependent Schrödinger equation of the aligned $D_2^+$ with the electric laser field is solved for the simulation of the complicated dissociative ionization processes and compared with the recent related experimental results. In this work, the R-dependent ionization rate and the enhanced ionization phenomenon beyond the Born-Oppenheimer approximation (BOA) are introduced and calculated. The substructure of the nuclear kinetic energy release spectra are revealed as the Coulomb explosion energy spectra and dissociation energy spectra in the dissociation-ionization channel. The significant and trace of these distinct sub-spectra in the total spectra comparatively are displayed and discussed.

## Introduction

$H_2$ and $H_2^+$, two basic molecules, are extensively studied experimentally and theoretically. These studies cause to appear comprehensive new phenomena [1]. Studies of the dynamic of $H_2$ ($D_2$) and $H_2^+$ ($D_2^+$) exposed to intense laser field are very complicated because involved two processes, ionization and dissociation simultaneously. In intense laser field, electron dynamics occur in attosecond time scales and nuclear dynamics, vibration and rotation, takes place in femtosecond and picosecond time scales. It is possible based on Born Oppenheimer approximation (BOA) to investigate these two dynamics, nuclear and electronic, separately. This approach is extensively used to investigate electronic dynamics of molecules in intense laser fields. When molecule is exposed to intense laser field, accurately probe of molecular dynamics that is involved simultaneously electronic and nuclear dynamics is very complicated. In these conditions, the perfect complicated simulation based on the solving of the time dependent Schrödinger equation (TDSE) beyond BOA, the most rigorous and adequate, ab-initio theoretical approach, would require to the complete description molecular dynamics. For a molecule with two or farther electrons, this task is very far from present available computer ability, even through without consideration the nuclear dynamics [2]. For the linear molecule with only one electron, this rigorous approach is feasible only for aligned molecule with the electric laser field together with the reduction of the dimensions of the electronic motion. Therefore, the most theoretical investigations were carried out for the dissociative ionization of the aligned $H_2^+$ ($D_2^+$) parallel to the electric field axis of the linear polarized laser pulse and three spatial coordinates of the electron in the TDSE are reduced to 1D based on an approximation known as quasi-Coulombic or soft-core (SC) Coulomb potential [3] to be able to carry out simulation, so that many research were done based on SC approximation [4-6] even for two electronic systems [7]. Nevertheless, the extent of ability and validity of SC approximation in the research of electron dynamics in multi-electron systems especially involved one and two electrons has been the subject controversy. The results of this work will explore this question. In this research, the previous studies are extended. We have done perfect complicated simulation of $D_2^+$ beyond BOA and also without SC approximation, i.e., by the rigorous solution of the TDSE for the full dimensional electron dynamics and also with consideration the nuclear dynamics of $D_2^+$ that is aligned with the electric laser field.

## Numerical solution of the TDSE and discussion

Time dependent Schrödinger equation in the cylindrical polar coordinate system for $H_2^+$ ($D_2^+$) located in the laser field as $E(t) = E_0 f(t)\cos(\omega t)$ parallel to the inter-nuclear axis in atomic unit ($\hbar = m_e = e = 1$) reads as (throughout of this article we use the atomic unit unless stated)

$$i\frac{\partial \psi(z,\rho,R,t)}{\partial t} = H(z,\rho,R,t)\,\psi(z,\rho,R,t), \tag{1}$$

where the total 3D electronic Hamiltonian is given by [8-10]

$$H(z,\rho,R,t) = -\frac{2m_N + m_e}{4m_N\, m_e}\left[\frac{\partial^2}{\partial\rho^2} + \frac{1}{\rho}\frac{\partial}{\partial\rho} + \frac{\partial^2}{\partial z^2}\right] - \frac{1}{m_N}\frac{\partial^2}{\partial R^2} + V_C(z,\rho,R,t) \tag{2}$$

$$V_C(z,\rho,R,t) = \frac{-1}{\sqrt{(z+R/2)^2+\rho^2}} + \frac{-1}{\sqrt{(z-R/2)^2+\rho^2}} + \frac{1}{R} + \left(\frac{2m_p + 2m_e}{2m_p + m_e}\right) z E_0\, f(t)\cos(\omega t) \tag{3}$$

with $E_0$ being the laser peak amplitude, $\omega = 2\pi\upsilon$ the angular frequency, and finally *f(t)* the laser pulse envelope which is set as

$$f(t) = \exp\left[\frac{-2\ln(2)\,(t-t_{on})^2}{\tau_p^2}\right], \qquad (4)$$

where $\tau_p$ is the full width at half maximum (FWHM) duration of the Gaussian shape of the pulse of laser. The laser pulse in this simulation suddenly turns on at time $t_{on}$. We are assuming $t_{on}$ is the time on which a $H_2^+$ ($D_2^+$) is suddenly created according a Frank-Condon transition from natural $H_2$ ($D_2$) onto $H_2^+$ ($D_2^+$) $\sigma_g$ state (Fig. 1).

In intense laser field, some parts of $D_2^+$ wavepackage become unbound and outgoing through different channels. A part of this unbound wavepacket becomes outgoing as $D+D^+$ through dissociation channel (DC) that is not concerned in this article and an other part becomes outgoing through dissociation-ionization channel (DIC) as $D^++D^+$ (Fig. 1) that we study in this article. The nuclear components in this channel (DIC) possess both dissociation energy (DE) and Coulomb explosion energy (CEE).

The accurate kinetic spectra of different decay channel, i.e. ionization and dissociation or both simultaneously, can be determined by the virtual detector method [10-11]. The virtual detector method makes it possible to precisely define and distinguish outgoing norms and energy from different DC and DIC channels. This method allows us to accurately determine CEE and DE of nuclear fragments through DIC. More details about the virtual detectors method and its abilities were represented in our previous reports [10,18].

One of the main purposes of this work is comparison between the present simulation results and the recent experimental results [5,6]. The experimental researches that up until now to be directly started with $H_2^+$ and $D_2^+$ molecular ion are extremely few [12]. Most experiments have been performed using $H_2$ and $D_2$ molecules and during the rise of the femtosecond laser pulse, $H_2^+$ ($D_2^+$) molecular ion are created. These pulses are usually focused to the peak intensities of ~$10^{13}$–$10^{15}$ W/cm$^2$ into the gas jet of unaligned $H_2$ or $D_2$ neutral molecules. When $H_2$ ($D_2$) molecule is exposed to a linearly polarized intense femtosecond laser pulse, the first electron is ejected during rising laser pulse. We assume that the ejection of this first electron is occurred at the time $t_{on}$ instantaneously via tunneling (Fig. 1). Also, we suppose that this process prepares a nuclear wave packet identical to the initial vibrational state of $H_2$ ($D_2$) via a vertical Franck-Condon transition onto the $H_2^+$ ground $\sigma_g$ state as depicted in Fig. 1. In addition, we assume that at the time of the ejection of the first electron, $H_2$ ($D_2$) molecule (and then $H_2^+$ ($D_2^+$)) is aligned with the linearly polarized intense femtosecond laser pulse. During the remained parts of the laser field envelope, the complicated dissociation-ionization processes of $H_2^+$ ($D_2^+$) take place.

In this work, $D_2^+$ is exposed to femtosecond laser pulses with the two different FWHM duration ($\tau_p$=40 and 140fs). The intensities of these two femtosecond laser pulses are equal to $1.0\times10^{14}$ W cm$^{-2}$ but their wavelengths are different and equal to 800 and 1200 nm respectively. In these simulations, the femtosecond laser pulses turn on suddenly at two cycles before the peak of the laser pulse envelop and the simulations start just at this moment. The nuclear kinetic energy release (KER) related to only the DIC for this molecular ion in these intense short laser pulses are showed in Figure 2. The results of this simulation (black lines) are compared with the experimental results (dotted lines) [5-6] and theoretical calculation with SC approximation (grey lines) [5-6]. In the experimental research [5-6], the gas jet of unaligned $D_2$ and $H_2$ exposed to the femtosecond laser pulses. It is import to note that the uncertainty in the peak intensity in the experimental work [5-6] is about 10% at 800 nm and 30% at 1.2 μm and 1.4 μm wavelength and the uncertainty in the pulse duration is about ±10%. It is surprising that in spite of the several considered assumption in the setup of simulation and the mentioned uncertainty involved in the experiment, there are good agreement between the KER of the experiment and simulation. There is an explicit different between theoretical and experimental results in the high KER above ~7.6 eV, in Fig. 2(a). This high KER in the experimental results seems to be related to the effect of nonaligned molecules (Fig. 16(c) of the Ref. [6]).

The KER of DIC ($D^{+}$+$D^{+}$) contains both simultaneously dissociation and ionization energies (DIE), i.e. it possesses CEE ($1/R$) and DE. The signification of the DE in the KER spectra is often ignored. We recently predicted that the DE spectra are not sharp and ignorable and their structure can cause considerable changes in the structure of the KER spectra [13]. The simulation results for the DE and CEE spectra are depicted in Fig. 3 with blue and grey lines respectively and compared with the KER of DIC (red lines) taken from Fig. 2. The Fig. 3 shows precisely that the DE spectra do not have a very sharp peak about a distinct value but a relatively wide Gaussian shape distributions that show peaks for 3(a) about ~1.6 eV and for 3(b) about ~0.3eV. The overall shapes of the KER spectra resemble to the CEE spectra. These figures show that it is possible to relate various parts of the KER and CEE spectra together. The numbers over graphs show these relations.

Fig. 3 shows that the overall shape of the KER and CEE spectra are similar. The extent of modulation and displacement of KER in respect of CEE depends on distribution of DE. The magnitude of the displacement of the peaks in Fig. 3(a) is about ~1.6, i.e. correspond to the place of peak of DE in Fig. 3(a). The average of the magnitude of the displacement in Fig. 3(b) is also about the place of the maximum of DE spectrum, i.e. ~0.3eV. Therefore, the magnitude of the displacement in Fig. 3(b) compared to 3(a) is small but the structure of the KER with respect of CEE in Fig. 3(b) is more complicated than Fig. 3(a) and more peaks appear in the KER spectrum of Fig. 3(b) with respect to Fig. 3(a). Therefore, comparison of the spectra in figures 3(a) and 3(b) show that altogether the DE in Fig 3(b) has a sharper Gauss shape distribution and results smaller displacements in the KER spectrum in comparison with CEE but results more complicated structure for KER in respect of CEE. The corresponding peaks of the KER and CE in Fig 3(a) can be related easier than Fig 3(b). The magnitude of the overall displacement of the KER spectra with respect to the CEE spectra helps us to determine these relations (the numbers on the graphs determine these relations).

An interesting and complicated phenomenon for $H_2^+$ in the intense laser field is the enhancement of the ionization rate as a function of inter-nuclear separation which results in maxima at some critical inter-nuclear distances ($R_C$). This phenomenon has been showed theoretically, but with BO approximation [13,14]. The $H_2^+$ molecular ion exhibits some critical distances ($R_C$) at which the molecular ionization rate exceeds the atomic rate by several order of magnitude and propose that the last electron is ejected mainly at much longer internuclear distances than the equilibrium internuclear separation. This effect has an special position in the interpretation of the molecular dynamic and fragmentation in the intense laser field [5-6,15] and Coulomb explosion imaging (CEI) [16]. Recently, the precise calculations of the ionization rates based on the BOA have been developed [13-14,17].

It is surprising and important to investigate the (enhanced) ionization rate and critical distance beyond BOA. It is clear that whenever $R_C$ becomes longer, the CEE and then total KER becomes smaller. When the electron of $D_2^+$ is ejected mostly at $R_C$, the two positive deuteron are exposed to mutually simultaneously repulsive force (Coulomb explosion). The magnitude of this repulsive force is proportion to the reciprocal of $R_C$. Therefore, it is predictable from the results of Figures 2 that the $R_C$ of the Fig. 2(b) to be longer than the Fig. 2(a). Fig. 4 shows the R-dependent nuclear distributions of the DIC of $D_2^+$ that respectively related to Fig. 2 and 3. Fig. 4 represents the real enhanced ionization beyond BOA. The R-dependent nuclear distributions and Kinetic spectra in figures 2-4 have been derived by the virtual detector method [10,11]. The structure and variations of KER in Fig. 2 can be related to the structure of the R-dependent non-BOA ionization rates in Fig. 4 as shown by numbers on the plots in the figures 3 and 4. The plots in Fig. 4 show the R-dependent non-scaled ionization rates.

The CEE spectra in Fig. 3 have been extracted by using of the results of Fig. 4. The relation of the various parts of Fig. 4 and the CEE spectra in Fig. 3 have been determined with the depicted numbers. Fig. 1 shows that the maximum amplitude of the nuclear distribution functions of $H_2^+$ created by the Frank-Condon transition from $H_2$ is about R~1.5 but Fig. 4 shows that the ionization rate for the points about this nuclear distance (~1.5) is negligible. The maximum of the ionization rates are placed for Fig. 4(a) about ~4.4 and for Fig. 4(b) about ~6. Fig. 5 of the Ref. [18] for I= $1\times10^{14}$ W/cm$^2$ intensity and $\lambda$ = 1064 nm wavelength shows that $R_c$'s are appear about ~6 and 9.5 for $\tau_1 = 0$ (the rising time of the laser pulse) and Fig. 1 of the Ref. [10] for I= $1\times10^{14}$ W/cm$^2$ intensity and $\lambda$ = 790 nm wavelength shows that $R_c$'s are placed about ~5, 7, and 9.5. Comparison of the non-BOA in Fig. 4 in the present study with the BOA in the references of [10,18] shows that some $R_C$'s appear in the both non-BOA and BOA. As we mentioned in a previous report [13], when the duration of ionization is short, the motion of nuclei are slow during the course of ionization process and the system has a little chance to evolve to large values of R during passing laser pulse, and therefore, the ionization signal for larger values of R will be weak or even null. Therefore, initial nuclear distribution and duration and intensities of the pulse laser are the effective parameters that determine the shape of ionization signal.

Comparison of the BOA and non-BOA shows that these parameters in the simulations in this work are so that results in EI to be strong only for small values of nuclear distance in the ionization signals.

## Conclusion

In summary, it appears that the electronic full dimensional TDSE without any SC approximation can simulate very well the recent experimental results and the Sc approximation can be success in some case and fails to simulate some experimental results. In simulation of one and two electron system often SC approximation are used. This work to some extent explores the validity and ability of this approximation and possibility of the application of SC approximation to study and simplify such complicated systems.

The relation of the nuclear kinetic energy distributions of the DIC and the R-dependent ionization rate clearly was explored and thus the EI phenomenon was confirmed beyond the BO approximation. Therefore, on the base of these simulations, we confirm directly the existence of the critical internuclear distance.

It was showed that in the KER structure, although the CEE substructure plays fundamental role but the DE substructure results in the relative KER spectrum is modulated and displaced with respect to the CEE spectrum. Therefore, in precise reconstruction of vibrational nuclear wave function or wave packet from the total KER spectra of DIC, the accurate determination of the structure of the both CEE and DIE spectra is necessary.

I have benefited from valuable and stimulating discussions with Prof. H. Sabzyan. The author would like to thank the University of Isfahan for financial supports, research facilities and also Isfahan High Performance Computing Center (IHPCC).

## Figure Captions:

**Fig. 1.** Sketch of the preparation and evaluation $H_2^+$ exposed to a linearly polarized intense femtosecond laser pulse. When $H_2$ molecule is exposed to the intense laser pulse, the first electron is ejected at time $t_{on}$ instantaneously. This process prepares a nuclear wave packet identical to the initial vibrational state of $H_2$ by a vertical Franck-Condon transition onto the $H_2^+$ ground $\sigma_g$ state. During the remained parts of the laser field envelope, the complicated simultaneously dissociation-ionization processes of $H_2^+$ take place. Some parts of $D_2^+$ wavepackage become unbound and outgoing through different channels, as $D+D^+$ through dissociation channel (DC) and as $D^+ +D^+$ through dissociation-ionization channel (DIC). The nuclear components in DIC possess both dissociation energy (DE) and Coulomb explosion energy (CEE).

**Fig. 2.** The kinetic energy release (KER) spectra of nuclear energy of $D_2^+$ exposed to femtosecond laser pulses with the different FWHM duration and wavelength: (a) $\tau_p$=40 of I= $1\times10^{14}$ W/cm$^2$ intensity and $\lambda$ = 800 nm wavelength and (b) 140fs of I= $1\times10^{14}$ W/cm$^2$ intensity and $\lambda$ = 1200 nm wavelength. The results of this work (black lines) are compared with the results of the experimental research (dotted lines) [7-8] and the theoretical calculation with SC potential approximation (grey lines) [7-8].

**Fig. 3.** KER spectra (red lines) and their substructures the CEE (grey lines) and DE spectra (blue lines).

**Fig. 4.** The nuclear R-dependent distribution of the ionization channel of $D_2^+$ for the two related simulations in Fig. 2 respectively.

**Figure 1**

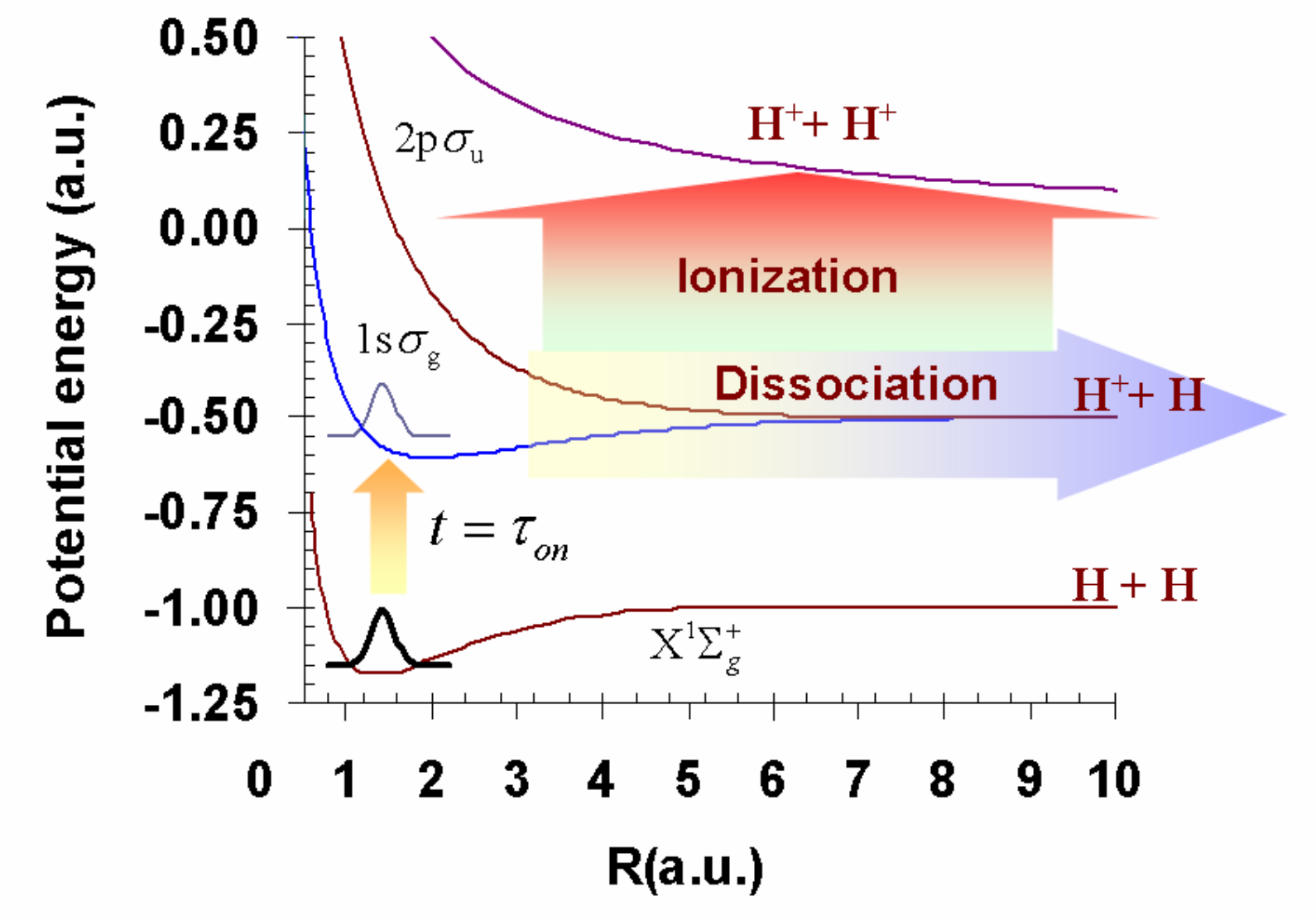

Potential energy (a.u.)
0.50
0.25
0.00
-0.25
-0.50
-0.75
-1.00
-1.25
0
1
2
3
4
5
6
7
8
9
10
R(a.u.)
$2p\sigma_u$
$1s\sigma_g$
$H^+ + H^+$
Ionization
Dissociation
$H^+ + H$
$t = \tau_{on}$
H + H
$X^1\Sigma_g^+$

**Figure 2**

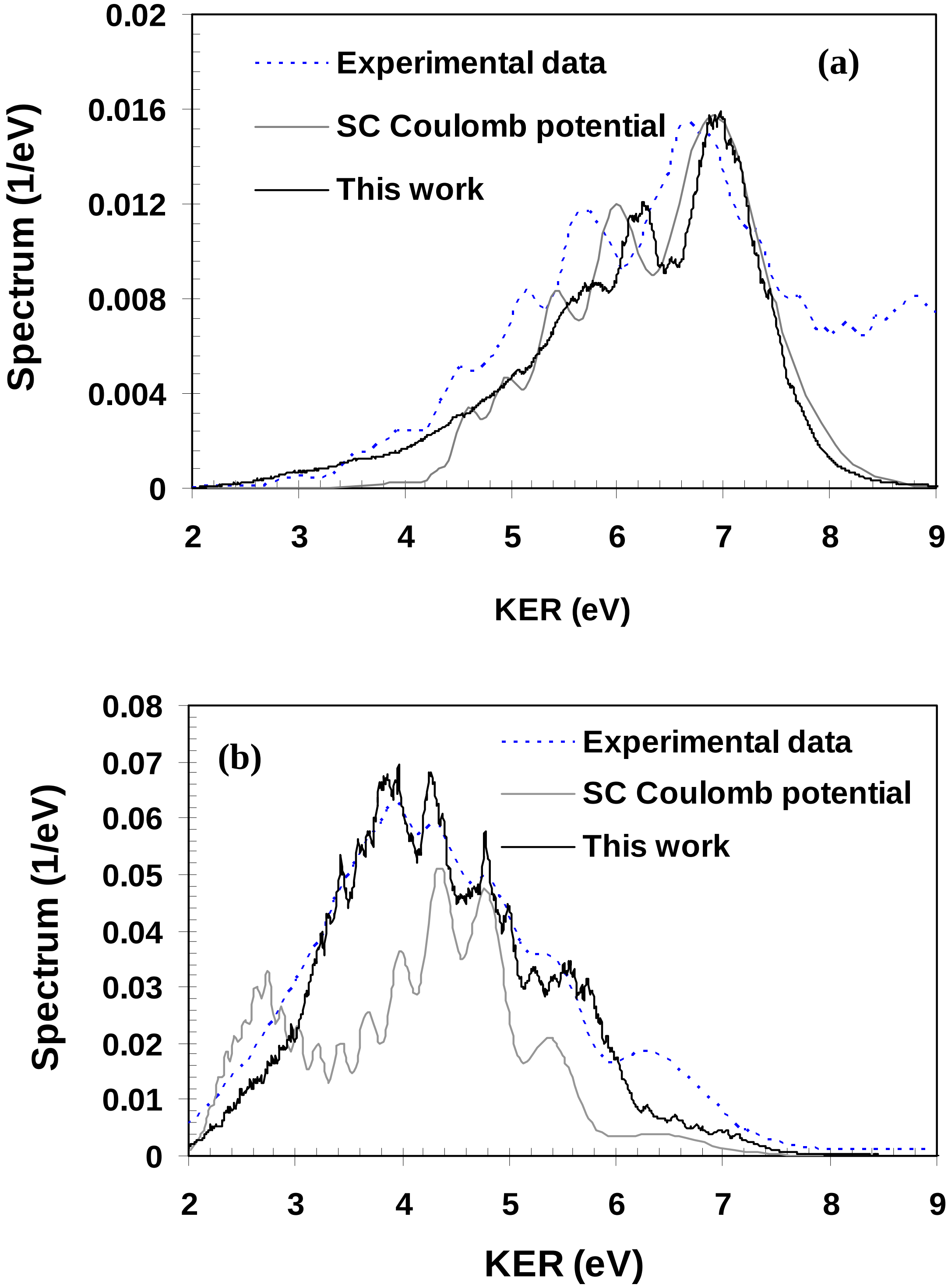

(a)
Experimental data
SC Coulomb potential
This work
Spectrum (1/eV)
0.02
0.016
0.012
0.008
0.004
0
2
3
4
5
6
7
8
9
KER (eV)
(b)
Experimental data
SC Coulomb potential
This work
Spectrum (1/eV)
0.08
0.07
0.06
0.05
0.04
0.03
0.02
0.01
0
2
3
4
5
6
7
8
9
KER (eV)

**Figure 3**

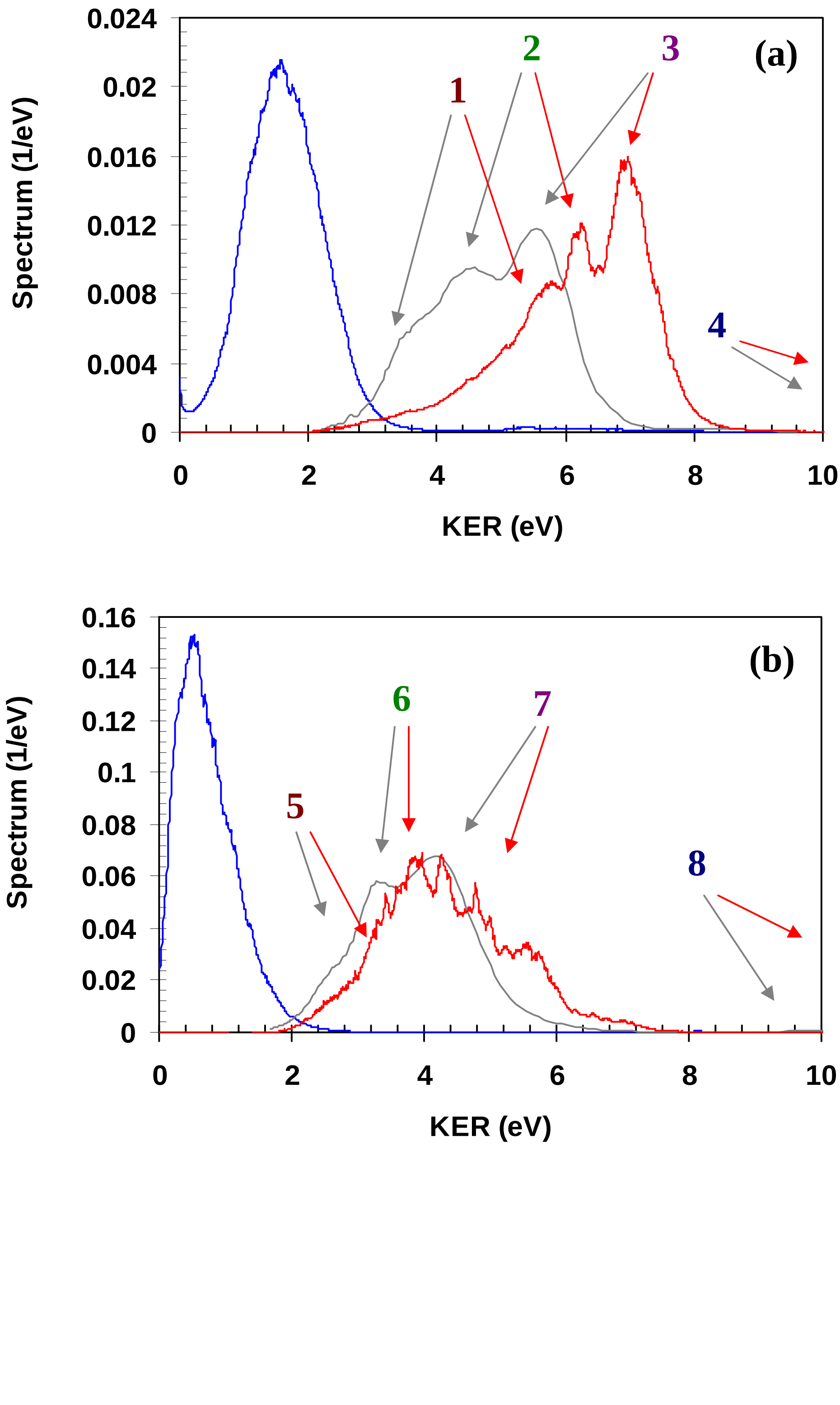
(a)
Spectrum (1/eV)
0.024
0.02
0.016
0.012
0.008
0.004
0
0
2
4
6
8
10
KER (eV)
1
2
3
4
(b)
Spectrum (1/eV)
0.16
0.14
0.12
0.1
0.08
0.06
0.04
0.02
0
0
2
4
6
8
10
KER (eV)
5
6
7
8

**Figure 4**

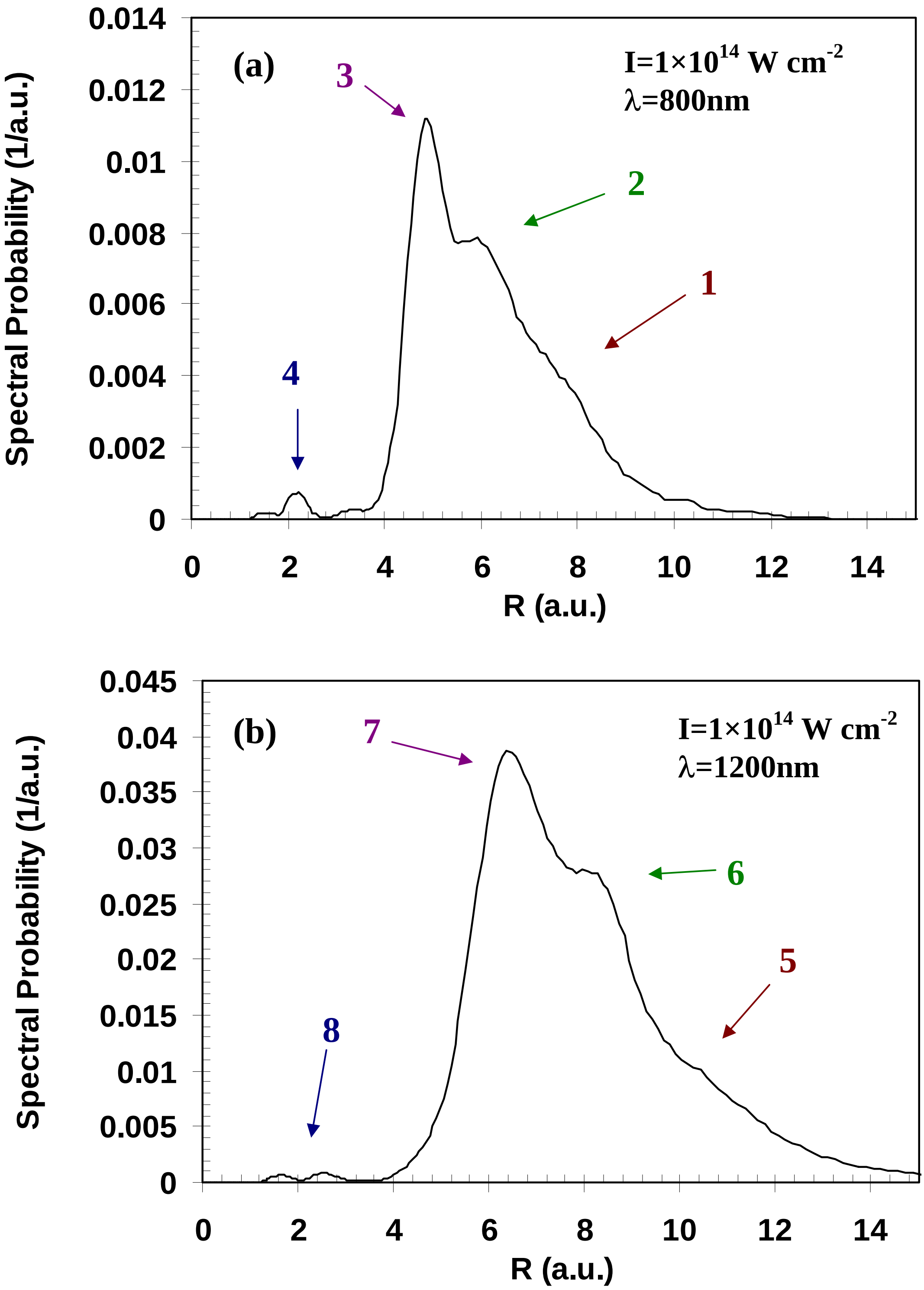
(a)
I=1×10^14 W cm^-2
λ=800nm
1
2
3
4
Spectral Probability (1/a.u.)
R (a.u.)
0.014
0.012
0.01
0.008
0.006
0.004
0.002
0
0
2
4
6
8
10
12
14
(b)
I=1×10^14 W cm^-2
λ=1200nm
5
6
7
8
Spectral Probability (1/a.u.)
R (a.u.)
0.045
0.04
0.035
0.03
0.025
0.02
0.015
0.01
0.005
0
0
2
4
6
8
10
12
14